\documentstyle[12pt]{article}
\begin{document}

\begin{center}

{\bf\large{Observation of  S=+1 narrow resonances in the system
$K^0_sp$ from p+$C_3H_8$ collision at 10 GeV/c} }

\vskip 5mm P.Zh.Aslanyan$^{a,b}$, V.N.Emelyanenko$^a$,
  G.G. Rikhkvitzkaya$^a$

\vskip 5mm {\small (a) {\it Joint Institute for Nuclear Research}}
\\
(b) {\it Yerevan State University }
\\
$\dag$ {\it E-mail: petros@spp.jinr.ru}
\end{center}

\vskip 5mm
% ----------------------------------------------------------------
\begin{abstract}
The 2m propane bubble chamber experimental data have been analyzed
to search for an exotic baryon state, the $\Theta^+$ -baryon, in
the $pK^0_s$ decay mode for the reaction p+$C_3H_8$ at 10 GeV/c.
The $pK^0_s$ invariant mass spectrum shows resonant structures
with $M_{K_s^0 p}$=1540$\pm$8, 1613$\pm$10, 1821$\pm$11 MeV/$c^2$
and $\Gamma_{K_s^0 p}$= 9.2$\pm$1.8, 16.1$\pm$4.1, 28.0$\pm$9.4
MeV/$c^2$. The maximal statistical significance of these peaks
have been estimated as 5.5$\sigma$,4.8$\sigma$ and 5.0$\sigma$,
respectively.
\end{abstract}

% ----------------------------------------------------------------
\section{Introduction}

Recent experimental efforts were largely motivated by Diakonov,
Petrov and Polyakov \cite{1} who studied antidecuplet baryons
using the chiral soliton (Skyrme) models. The lightest member of
the pentaquarks antidecuplet,$\Theta^+$(1530) -baryon predicted in
\cite{1}, has positive strangeness, the mass of M=1530 MeV/c$^2$,
1/2  spin  and even parity. Jaffe and Wilczek have suggested an
underlying quark model structure of this state \cite{2}. There are
other theoretical simulations which have predicted for this state
\cite{3,4,5}. The rotational states of the S=+1 $\Theta^+$ baryons
are shown in the paper, by David Akers\cite{5} . Experimental
evidence for $\Theta^+$(1530)-baryon with positive strangeness has
come recently from several experimental groups (LEPS\cite{6}],
DIANA-ITEP \cite{7}, CLAS \cite{8}, SAPHIR \cite{9}, HERMES
\cite{10}, SVD-2 experiment IHEP \cite{11}. The reaction CC$\to
pK^0_s$ +X with the propane bubble chamber technique has observed
a resonant structure in the p$K^0_s$ invariant mass spectrum with
$\Theta^+$(1532) \cite{12}.

\section{ Method}

A reliable identification of the above mentioned resonance needs
to use 4$\pi$-detectors and high precision measurements of the
sought objects.  The 2 m propane bubble chamber is the most
suitable instrument for this purpose \cite{13,14}. The
experimental information of more than 2*350000 stereo photographs
are used to select the events with $V^0$ strange particles . The
based on the program GEOFIT\cite{15} is used to measure  the
kinematics parameter  of tracks - p, $tg\alpha$, $\beta$ in the
photographs. The relative error of measuring momentum p and the
average length L of charged particles are\\
$<\Delta p/p>$=2.1\%, $<L>$=12 cm for stopping particles;\\
$<\Delta p/p>$=9.8\%, $<L>$= 36 cm for not stopping particles.\\
The mean values of measurement errors for the dip and azimuthal
angles are relatively equal to $<\Delta tg\alpha> = 0.0099\pm
0.0002$ and $<\Delta\beta> = 0.0052\pm 0.0001$ (rad.).

The estimate of ionization, the peculiarities of the end track
points of the stopping particles(protons,$K^{\pm}$) allowed one to
identify them. Protons could be identified over the following
momentum range: $0.150\le p\le $ 0.900 GeV/c. In the second
momentum range $p> $ 0.900 GeV/c protons couldn't be separated
from other particles. Therefore, the experimental information has
been analyzed similarly two separate ranges.

\subsection{Identification of $\Lambda$ and $K^0_s$ }

The events with $V^0$ ($\Lambda$ and $K^0_s$)  were identified
using the following criteria \cite{16,17}:\\
 1)two-prong stars from the photographs were selected
according to $\Lambda\to\pi^-+p$, neutral $K_s\to\pi^-+\pi^+$ or
$\gamma \to e^++e^-$ hypothesis ; 2) $V^0$ stars should have the
effective mass of $K^0_s$ and $\Lambda$; 3) these $V^0$ events are
directed to  some vertices(complanarity); 4) they should have one
vertex a three constraint fit for the hypothesis with $M_K$,
$M_{\Lambda}$ at $\chi^2 <12$ and a momentum limit of $K^0_s$ and
$\Lambda$ greater than 0.1 and 0.2 GeV/c, respectively; 5)The
analysis has shown\cite{17} that the events with undivided
$\Lambda K^0_s$ were appropriated events as $\Lambda$.

Figures.1a,c and 1b,d show the effective mass distribution of
$\Lambda$(8657-events),$K^0$(4122-events) particles and  their
$\chi^2$ from kinematics fits, respectively, produced from the
beam protons interacting with propane targets. The measured masses
of these events have the following Gaussian distribution
parameters $<M(K_s)>$= 497.7$\pm $3.6, s.d.= 23.9 MeV/$c^2$ and
$<M(\Lambda)>$ =1117.0$\pm$ 0.6, s.d.=10.0 MeV/$c^2$. The masses
of the observed $\Lambda$, $K^0_s$ are consistent with their PDG
values \cite{18}. The effective mass of the $\Theta^+\to K^0p$
system, like that of the $\Lambda \to \pi^- p$ system, is measured
to a precision of $<\Delta M_{(K^0_sp)}/M_{(K^0_sp)}> \approx$
1.1\%. Then the effective mass resolution of $K^0p$ system  was
estimated to be on the average 0.6\% for identified protons with a
momentum of $0.150\le p\le $ 0.900 GeV/c.

 The preliminary estimate of the experimental total cross
sections for $K^0_s$ production in the p$^{12}C$ collisions at 10
GeV/c is equal to $\sigma$= 3.8$\pm$0.6 mb.

\section{$pK^0_s$ - spectrum  analysis}

\subsection{$pK^0_s$ - spectrum for identified protons with a momentum of
$0.350\le p_p\le 0.900$ GeV/c }

The $pK^0_s$ effective mass distribution for 2300  combination is
shown on Fig.2. The solid curve is the sum of the background and 4
Breit-Wigner resonance curves.

The total experimental background has been  obtained by two
methods. In the first method, the experimental effective mass
distribution was approximated by the polynomial function  after
cutting out the resonance ranges because this procedure has to
provide the fit with $\chi^2$=1 and polynomial coefficient with
errors less than 10\%. This distribution was fitted by the eight-
order polynomial. The second of the randomly mixing method of the
angle between  $K^0_s$ and p for experimental events is presented
in \cite{21}. Then, these background events were analyzed by using
the same experimental condition and the effective mass
distribution $pK^0_s$ was fitted by the eight-order polynomial.
The analysis done by two methods has shown that while fitting had
these distributions the same coefficient and order of polynomial.

 The background for $\overline{K^0}p$ combinations is
estimated with FRITIOF model \cite{19,20} and no more than 10\%
has been obtained. No obvious structure in $\overline{K^0}p$
spectrum is seen  in Fig.2.

 The statistical significance for the fit in Fig. 2 is
calculated as NP /$\sqrt{NB}$, where NP is the number of events
corresponding to the signal on top of the fitted background and NB
is the number of events corresponding to the background in the
chosen area. There are significant enhancements in 1540,1612 and
1821 MeV/$c^2$ mass regions. There are small peaks in 1480 and
1980 MeV/$c^2$ mass regions.

\subsection{$pK^0_s$ - spectrum for positively charged, relativistic tracks
with a momentum of $0.9 \le p_p\le 1.7$ GeV/c }

 The $pK^0_s$ invariant
mass spectrum shows  resonant structureû with M = 1515
(5.3$\sigma$) and 1690 MeV/$c^2$(3.8$\sigma$) in Fig.3a. No
obvious structure in 1540,1610 and 1821 MeV/$c^2$ mass regions is
seen in Fig3a . FRITIOF \cite{19,20} model shows that the average
multiplicity in this range for: all positive tracks, protons and
$\pi^+$ are equal to 1.2, 0.4 and 0.8, respectively. The
background for $K^0_s\pi^+$ and $K^0_s K^+$ combinations are equal
to $\approx$ 46.6\% and 4.4\% respectively. These observed peaks
can  be a reflection from resonances $\Lambda(1520)$ and
$\Lambda(1700)$. The (n$\overline{K^0}$ ) invariant mass spectra
for events where $\pi^+$ -meson in reactions p+$C_3H_8\to \pi^+
\overline{K^0} nX$ was detected and its mass was substituted by
the mass of neutron.

\subsection{$pK^0_s$ - spectrum for positively charged tracks
 with a momentum of $p_p\ge 1.7$ GeV/c}

The $pK^0_s$ effective mass distribution with a momentum $p_p\ge $
1.7 GeV/c(3500 combination) is shown in Fig.3b. The histogram is
approximated by a polynomial background curve and by 5 resonance
curves taken in the Breit-Wigner form. The dashed curve is the
background taken in the form of a superposition of Legendre
polynomials up to the 6 -th degree, inclusive. The analysis done
by two methods has shown that while fitting had these
distributions the same coefficient and order of polynomial. The
average multiplicity(FRITIOF) in this range for: all positive
tracks, only protons and only $\pi^+$ are equal to 1.3, 0.8 and
0.5, respectively. Therefore the background for $K^0_s\pi^+$ and
$K^0_s K^+$ combinations are equal to $\approx$ 20\% and 5\%,
respectively. The estimate of contribution for $\overline{K^0}p$
combinations with FRITIOF model is equal to 9\%. No obvious
structure in $\overline{K^0}p$ spectrum is seen on fig.3b.

 There are significant enhancements in 1487, 1544, 1612 and
1805 MeV/$c^2$ mass regions. Their excess above background is 3.0,
3.9, 3.7 and 4.0 S.D., respectively. There is a small peak in
1693 MeV/$c^2$ mass regions.
%----------------------------------------------------------------

\subsection{The sum of $pK^0_s$ - spectrum for identified protons
and  positively charged tracks($p_p\ge 1.7$ GeV/c)}

The total effective mass distribution for $pK^0_s$(5554
combination) is shown on Fig.4. The solid curve is the sum of the
background and 4 Breit-Wigner resonance curves. The background was
fitted by the six-order polynomial. The total experimental
background(dashed histogram)   with the same experimental
condition is also obtained  by the second method \cite{21}. The
dashed curve is the background taken in the form of a
superposition of Legendre polynomials up to the 6 -th degree,
inclusive.

There are significant enhancements in 1545, 1616,1690,1811 and
1980 MeV/$c^2$ mass regions. Their excess above background is by
the first method 6.2,6.6, 4.0, 5.0, 3.0 S.D., respectively and by
the second method 5.5, 4.8, 3.6, 5.0,3.0 S.D., respectively.

\section{Conclusion}

The effective mass spectra $K^0_s p$ in collisions protons of a
10.0 GeV/c momentum with $C_3H_8$ nuclei, have led to the
discovery of the peaks presented below(Table 1). Table 1 shows the
width($\Gamma$) and the effective mass resonances  which are
based on the data from Fig. 2. The statistical
 significance are the based on the data from Fig.4.
 The primary total cross section for $\Theta^+{(1540)}$
 production in $pC_3H_8$-interactions is estimated
 to be $\approx~90 \mu$b.

\newpage
\vskip 2mm
\begin{center}
\Large Table 1.\\
\end{center}
\begin{tabular}{|c|c|c|c|c|c|c|c|}  \hline
Resonance &M&$\Gamma_e$&$\Gamma$&significance \\
system& MeV/$c^2$&MeV/$c^2$&MeV/$c^2$&(maximal)\\
 &&Experiment&&$N_{sd}$\\ \hline
$K^0_sp$&1540$\pm$8&18.2$\pm$2.1&9.2$\pm$1.8&5.5$\pm$0.5\\
$K^0_sp$&1613$\pm$10&23.6$\pm$6.0&16.1$\pm$4.1&4.8$\pm$0.5\\
$K^0_sp$&1821$\pm$11&35.9$\pm$12.0&28.0$\pm$9.4&5.0$\pm$0.6\\
\hline
\end{tabular}
%\end{table}

 \section{Acknowledgements}

The authors are very much obliged to  A.N. Sisakian, A.I.Malakhov,
Yu.A. Panebratsev, H.A. Vartapetian  for their support of this
work.

We are also greatly indebted to our colleagues A.A. Kuznetsov,
 E.N.Kladnitskaya,Yu.A.Troyan, M.V. Tokarev, A.H.Khudaverdyan,
A.S.Danagulian, V.I. Moroz, V.V. Uzhinskii  for the assistance in
data processing and fruitful discussions.

%\begin{figure}[htb]
%\centerline{\epsfig{file=fig4mr.eps,width=8.0cm}} \caption{figure
%caption. }
%\end{figure}

%\begin{figure}
%\centerline{\epsfxsize=4.4in\epsfbox{fig4mr.eps}}
%\caption{}
%\begin{picture}(300,450)
%\put (0,400) {\special{em:graph fig4mr.pcx}}
%w\end{picture}
%\end{figure}

\end{document}